\documentclass[12pt,a4paper]{article}

\usepackage{amssymb,amsthm,euscript,amsmath,graphicx,array}

\newcommand{\be}{\begin{equation}}
\newcommand{\ee}{\end{equation}}

\begin{document}
\title{Weakly nonlocal continuum theories of granular media: restrictions
from the {S}econd {L}aw}
\author{P. V\'an\\
{\small Budapest University of Technology and Economics}\\
{\small Department of Chemical Physics}\\
{\small 1521 Budapest, Budafoki \'ut 8}\\
{\small Email: vpet\@@phyndi.fke.bme.hu}}

\maketitle
\begin{abstract}
The classical continuum mechanical model of granular media of rational
thermodynamics results in a Coulomb-Mohr type equilibrium stress-strain
relation. The proof is based on a two component material model
introducing a scalar internal variable. Here we will show how one can
get similar stress-strain relations without assuming a balance of
substructural interactions, considering only the restrictions of the
Second Law of thermodynamics.
\end{abstract}

\date{ }

\maketitle


\section{Introduction}

In their classical paper Goodman and Cowin derived a material model of
porous and granular media using pure thermodynamic reasoning
\cite{GooCow72a}. They considered a material where the density of the
solid component is $\gamma$, the total density is $\rho$ and the {\em
volume distribution function} $\nu$ is defined by the following formula
$$\rho = \nu\gamma.$$
Later the scalar internal variable $\nu$ was interpreted as {\em
roughness} and its (substantial) time derivative $\dot{\nu}$ as {\em
abrasion} \cite{Kir02a,KirTeu02a}. Goodman and Cowin assumed a material
with incompressible solid component and with a balance form dynamic
equation for the abrasion, a balance of substructural interactions
\cite{Cap89b,Mar02a}:
$$
\gamma \ddot{\nu} = \nabla\cdot{\bf h} + \sigma_{\dot{\nu}},
$$

\noindent where ${\bf h}$ is the conductive current and
$\sigma_{\dot{\nu}}$ is the production of the abrasion. After
introducing a suitable constitutive space they investigated the
requirements coming from the Second Law of thermodynamics with
Coleman-Noll procedure. Their final result was a definition of {\em
Coulomb granular material} by the following stress function
$$
{\bf T}^e = (\beta_0 - \beta \nu^2 + \alpha \nabla\nu\cdot\nabla\nu +
2\alpha\nu\triangle\nu) {\bf I} - 2\alpha\nabla \nu \circ \nabla\nu +
\lambda Tr(\boldsymbol{\epsilon}){\bf I} + 2\mu\boldsymbol{\epsilon}.
$$

\noindent where $\beta_0, \beta,\alpha$ are material parameters,
$\lambda,\mu$ are the Lam\'e coefficients.  {\bf I} is the second order
unit tensor, $\circ$ denotes the tensorial product and $Tr$ is the
trace. As one can see an ideal elastic behavior is coupled to the
gradient dependent characteristic part, represented by the first two
terms.

In this paper we will show, that a similar material model can be
derived without assuming a balance of the abrasion, with the very same
weakly nonlocal extension of the configurational space as in the model
of Goodman and Cowin. In the following Liu's theorem will play an
important technical role. Details of the different state spaces, more
detailed description of thermodynamic concepts, the applied
mathematical methods (especially Liu procedure) can be found in
\cite{MusAta01a}, regarding the weakly nonlocal extension see
\cite{Van03a1,Van02m2}. A thermodynamic background of continuum field
theories is in \cite{Ver97b}.

\section{Weakly nonlocal fluids - granular media}

In our treatment the {\em basic state space} of granular media is
spanned by the density of the solid component, the volume distribution
function and the velocity ($\gamma, \nu, {\bf v}$). This basic state
space is the simplest large deformation treatment and considers the
possibility of changes in the topological structure (changes of
neighbours and rotation of grains) typical in fluids. With this basic
state we are constructing a constitutive model of a {\em dilatant
granular material}, because $\rho$ is not necessarily constant. The
{\em constitutive state space} contains gradients of the basic state
variables as in case of classical fluids. Therefore, it is spanned by
the variables $(\gamma, \nabla\gamma, \nu, \nabla\nu, {\bf v},
\nabla{\bf v})$. The functions interpreted on the constitutive space
are the {\em constitutive functions}. The {\em space of independent
variables} is spanned by the next time and space derivatives of the
constitutive variables $(\dot{\gamma}, \nabla\dot{\gamma},
\nabla^2\gamma, \dot{\nu}, \nabla\dot{\nu}, \nabla^2\nu, \dot{\bf v},
\nabla\dot{\bf v}, \nabla^2{\bf v})$, as a consequence of the entropy
inequality. Here $\nabla^2$ denotes the second space derivative.

In the following we will assume that the solid part of the material is
nearly incompressible, therefore the deformation is due to the changes
in the volume distribution function. This condition can be expressed as
\be \dot{\gamma} = 0 \label{inc_cond}\ee

Considering (\ref{inc_cond}) the continuity equation can be written as
\be \dot{\nu} + \nu \nabla\cdot{\bf v} = 0 \label{cont_cond}\ee

(\ref{inc_cond}) and (\ref{cont_cond}) are constraints of the
independent variables, and are to be considered in the application of
the Liu procedure. Moreover, the structure of the constitutive state
space implies that the space derivatives of the above equations contain
terms solely from the space of independent variables, therefore they
are also constraints. The derivative of (\ref{inc_cond}) is
 \be \nabla\dot{\gamma} = {\bf 0}, \label{dinc_cond}\ee

\noindent and the derivative of (\ref{cont_cond}) results in \be
\nabla\dot{\nu} + \nabla\nu \nabla\cdot{\bf v} + \nu \nabla
\nabla\cdot{\bf v} = 0. \label{dcont_cond}\ee

Finally, the balance of momentum is written as \be \gamma\nu \dot{\bf
v} + \nabla\cdot{\bf P} = 0 \label{mom_cond}.\ee

Here {\bf P} is the pressure tensor. The requirement of nonnegativity
of the entropy production is \be \gamma\nu \dot{s} + \nabla\cdot{\bf
j}_s =\sigma_s \geq 0 \label{Entr_bal},\ee

\noindent where the specific entropy $s$, the conductive current of the
entropy ${\bf j}_s$ and the pressure {\bf P} are the constitutive
quantities, functions interpreted on the constitutive space. With given
constitutive functions the dynamic equations of the granular continua
are (\ref{inc_cond}), (\ref{cont_cond}) and (\ref{mom_cond}). According
to the Second Law we are to find these constitutive functions that the
entropy production be nonnegative. In this way the nonnegativity will
be a pure material property, independently of the initial conditions.
Liu procedure is applied with the multiplier form (see
\cite{Liu72a,Van02m2})
\begin{eqnarray*}
\rho \dot{s} &+& \nabla \cdot {\bf j}_s -
    \Gamma_1 \dot{\gamma} -
    \Gamma_2 \nabla\dot{\gamma} -
    \Gamma_3 (\dot{\nu} + \nu \nabla\cdot{\bf v}) \\
    &-& \Gamma_4(\nabla\dot{\nu} +
        \nabla\nu \nabla\cdot{\bf v} + \nu \nabla \nabla\cdot{\bf v}) -
    \Gamma_5(\gamma\nu \dot{\bf v} + \nabla\cdot{\bf P})  \geq 0.
\end{eqnarray*}

Here we introduced Lagrange-Farkas multipliers $\Gamma_1, \Gamma_2,
\Gamma_3, \Gamma_4$ and $\Gamma_5$ for the constraints
(\ref{inc_cond}), (\ref{dinc_cond}), (\ref{cont_cond}),
(\ref{dcont_cond}) and (\ref{mom_cond}) respectively. In deriving the
Liu equations one should consider that the substantial time derivative
does not commute with the space derivative. The following identity is
to be applied
$$
\dot{(\nabla a)} = \nabla\dot{a} - \nabla{\bf v}\cdot \nabla a.
$$

Now the multipliers of the independent variables give the Liu
equations, respectively. Introducing a shorthand notation for the
partial derivatives of the constitutive quantities as e.g.
$\partial_\gamma = \frac{\partial }{\partial \gamma}$ we get
\begin{eqnarray}
\rho\partial_\gamma s &=& \Gamma_1, \label{L1}\\
\rho\partial_{\nabla\gamma}s &=& \Gamma_2, \label{L2}\\
\rho\partial_\nu s &=& \Gamma_3, \label{L3}\\
\rho\partial_{\nabla\nu}s &=& \Gamma_4, \label{L4}\\
\rho\partial_{\bf v} s &=& \rho\Gamma_5, \label{L5}\\
\rho\partial_{\nabla{\bf v}} s &=& {\bf 0}, \label{L6}\\
(\partial_{\nabla\nu}{\bf j}_s &-&
    \Gamma_5\partial_{\nabla\nu}{\bf P})^s = {\bf 0}, \label{L7}\\
(\partial_{\nabla\gamma}{\bf j}_s &-&
    \Gamma_5\cdot\partial_{\nabla\gamma}{\bf P})^s = {\bf 0}, \label{L8}\\
(\partial_{\nabla{\bf v}} {\bf j}_s &-&
    \Gamma_4\nu {\bf I} - \Gamma_5\cdot\partial_{\nabla{\bf v}} {\bf
    P})^s = {\bf 0}. \label{L9}
\end{eqnarray}

The superscript $ ^s$ denotes the symmetric part of the corresponding
function. Equations (\ref{L1})-(\ref{L5}) determine the Lagrange-Farkas
multipliers. The solution of (\ref{L6}) results in an entropy that is
independent of the gradient of velocity, therefore (\ref{L9}) can be
integrated and a particular form  of ${\bf j}_s$ is determined.
Substituting that form of ${\bf j}_s$ into (\ref{L7})-(\ref{L8}) we get
two equations that are fulfilled if
 \be \partial_{\nu\nabla\nu}s = {\bf 0}, \quad
\partial_{\nu\nabla\gamma}s = {\bf 0}, \quad
\partial_{\nabla\nu\nabla\nu}s = {\bf 0}, \quad
\partial_{\nabla\nu\nabla\gamma}s = {\bf 0}
\label{condL}\ee

The most general generalized entropy function, isotropic and second
order in {\bf v} and in $\nabla\gamma$ that satisfies the above
conditions is the following
 \be s(\nu,\nabla\nu,\gamma,\nabla\gamma,{\bf v}) =
    s_e(\nu,\gamma) - m(\nu,\gamma)\frac{{\bf v}^2}{2} -
    \alpha(\nu,\gamma)\frac{(\nabla\gamma)^2}{2}.
\label{ent_fun}\ee

Here $m$ and $\alpha$ are arbitrary nonnegative functions. We can see,
that entropy function is a concave function of the variables ${\bf v}$
and $\nabla\gamma$. Moreover, the entropy is independent of
$\nabla\nu$. Considering (\ref{ent_fun}), the solution of the last Liu
equation (\ref{L9}) gives the entropy current as
 \be {\bf j}_s = -m {\bf v}\cdot{\bf P} + {\bf j}_1(\nu,\gamma,{\bf v}).
\label{scur_fun}\ee

Here ${\bf j}_1$ is an arbitrary function. Applying (\ref{ent_fun}) and
(\ref{scur_fun}) the dissipation inequality simplifies to
$$
\nabla\cdot{\bf j}_1 - \nabla(m{\bf v}):{\bf P} + \rho(\alpha
\nabla\gamma \circ\nabla\gamma - \nu\partial_\nu s{\bf I}):\nabla{\bf
v} \geq 0
$$

If ${\bf j}_1\equiv {\bf 0}$ and $m=1$ (as usual) then the dissipation
inequality can be transformed into a solvable form. Introducing the
notation $p = -\nu\rho\partial_\nu s_e$ entropy inequality further
simplifies to
$$
- \left( {\bf P} - (p + \rho \partial_\nu\alpha
\frac{(\nabla\gamma)^2}{2}) {\bf I} - \rho\alpha
\nabla\gamma\nabla\gamma \right) :\nabla{\bf v} \geq 0
$$

The notation is not arbitrary, because $\partial_\nu s_e =
\partial_\rho s_e \partial_\nu \rho = -\frac{\tilde{p}(\nu,\gamma)}{T_0
\rho^2}$, where $\tilde{p}$ is the scalar pressure according to the
traditional thermostatic definition, corresponding to the Gibbs
relation. Or, alternatively investigating a pure mechanical system one
can recognize that our $s$ function is still connected to the entropy
only in some properties, therefore one can introduce the physical
entropy $\tilde{s}=\frac{s}{T_0}$. In this case $p$ is the scalar
pressure, $T_0$ is a constant temperature and all the considerations
above are valid. Let us underline here, that there is nothing
mysterious in the above identifications that would weaken the above
reasoning. Our whole procedure is based on the existence of a
constitutive function with the nonnegative balance, a kind of general
stability requirement. One can exploit this property and recognize the
physical meaning of the corresponding quantities later. On the other
hand we should not forget that the introduced $\tilde{s}$ is a kind of
generalized, nonequilibrium  "physical" entropy with rather strange
variables (velocity and gradients) and there is no unique nomination
for such quantities (coarse grained kinetic potential?).

The inequality above contains only the pressure tensor {\bf P} as
constitutive quantity that depends on the velocity gradient, therefore
it is solvable. The general solution is
$$
{\bf P} - {\bf P}_e = {\bf P}_v = -{\bf L}(\nabla{\bf v}),
$$

\noindent where ${\bf P}_v$ is the viscous part of the pressure, {\bf
L} is a nonnegative constitutive function. Here we assumed a symmetric
pressure tensor as it is usual for materials without internal moment of
momentum. Furthermore, the equilibrium, static part of the pressure is
${\bf P}_e$ and it is defined as \be {\bf P}_e = (p + \rho
\partial_\nu\alpha \frac{(\nabla\gamma)^2}{2}) {\bf I} - \rho\alpha
\nabla\gamma\nabla\gamma \label{CMpre_fun}\ee

A material defined by this equilibrium pressure tensor will be called
{\em Coulomb-Mohr material}.

\section{Coulomb-Mohr materials}

The pressure tensor (\ref{CMpre_fun}) has several remarkable
properties. First of all, the static shear stress is not zero, and
$\nabla\gamma$ is an eigenvector of ${\bf P}_e$. Furthermore,
introducing an arbitrary direction with a unit vector ${\bf n}$, one
can define a pressure normal to the direction ${\bf n}$ as
$$ N := {\bf n}\cdot{\bf P}_e\cdot{\bf n} =
\hat{p} - \rho\alpha(\nabla\gamma \cdot {\bf n})^2,$$

\noindent where $\hat{p} = p + \rho \partial_\nu\alpha
(\nabla\gamma)^2/2$. Denoting the shear pressure by $S:=P-N$ one can
get
$$
S^2+N^2 = ({\bf P}_e\cdot{\bf n})\cdot({\bf P}_e\cdot{\bf n}) =
\hat{p}^2 - 4\rho\alpha\hat{p}(\nabla\gamma\cdot{\bf n})^2 +
4(\rho\alpha)^2 (\nabla\gamma)^2(\nabla\gamma\cdot{\bf n})^2
$$

After a short calculations follows, that
$$S^2+(N-t)^2 = r^2 \quad \text{where} \quad
t=\hat{p}-r \quad\text{and}\quad r=\rho\alpha(\nabla\gamma)^2.$$

Hence, the possible Mohr circles in the material have a special
structure, their envelope from above is a straight line, the material
satisfies a failure criteria of Coulomb-Mohr kind, as the Coulomb
material of Goodman and Cowin. One can check that this is really a
failure criteria, the second derivative of the entropy function
(\ref{ent_fun}) become semidefinite on the Coulomb-Mohr line and the
line separates the regions of the state space where the thermodynamic
stability is violated from the region where it is fulfilled.

\begin{figure}[ht]
\centering
\includegraphics[height=7.5cm]{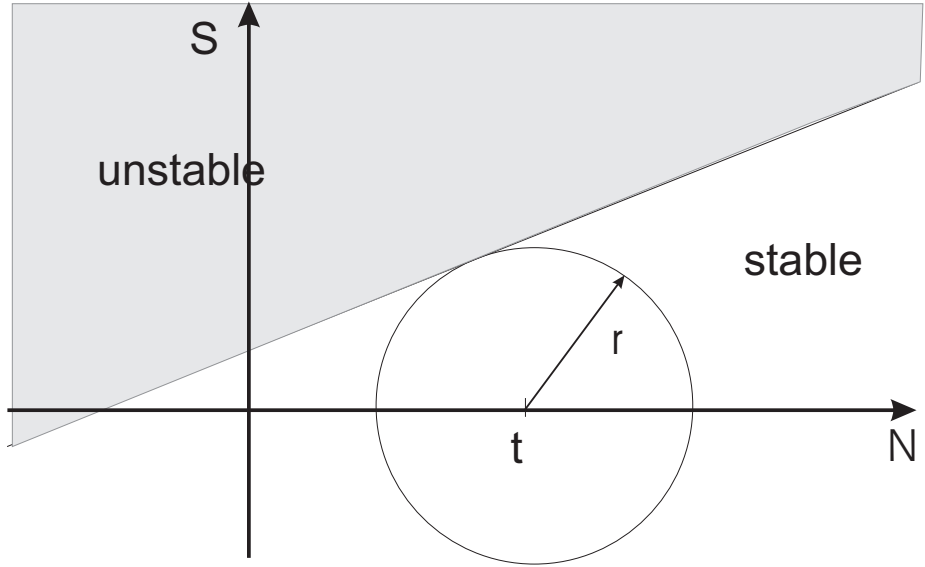}
\caption{The envelope of the Coulomb-Mohr circles as a boundary of the
stability} \label{fig2}
\end{figure}

Material stability in mechanics is far more complex matter than in
equilibrium thermodynamics of gases and fluids where it is connected to
the appearance of new phases \cite{God02a,God03a}. On the other hand in
a thermodynamic approach the loss of thermodynamic stability means the
violation of dynamic stability (e.g. the corresponding partial
differential equations loose their hyperbolicity). Stability losses of
this kind where treated in connection of rigid, microcracked materials
in \cite{Van01a1}. There the microstructure was characterized by a
vectorial dynamic variable {\bf D} (representing the average crack
length, a damage) and two general free energies of the microcracked
materials were introduced. There with special material parameters one
can arrive to a Coulomb-Mohr failure criteria and pressure tensors
similar to (\ref{CMpre_fun}), but with a local variable called damage
vector {\bf D}, instead of the gradient variable $\nabla\gamma$. The
role of $\nabla\gamma$ is similar to ${\bf D}$ in every sense.
Recalling the quadratic form of the entropy function the loss of
thermodynamic stability is not a direct consequence of an increasing
$|\nabla\gamma|$. The loss of stability is more involved and always
connected to changes in the stress/deformation state and appears only
in perpendicular to $\nabla\gamma$, in a 'shear' direction. The
complexity of the different kind stability losses is investigated e.g.
in \cite{Bed99a,Bed00a}.

Our basic state space and configuration space was that of a fluid. In
the one component case, when the configuration state space is
$(\rho,\nabla\rho,{\bf v}, \nabla{\bf v})$ an ideal Euler fluid and the
viscous Navier-Stokes fluids appear after similar derivations.
Remarkable, that a higher order weakly nonlocal one component fluid,
where the constitutive state space contains also the second order
derivative of the density, results in the Madelung fluid (known from
the hydrodynamic version of quantum mechanics) in the ideal, non
dissipative case \cite{VanFul03m}.

According to the comparison of Kirchner and Hutter the model of Goodman
and Cowin is the most robust one among several local continuum granular
material models of porous and granular media (with scalar internal
variables). Interestingly, the other compared models are all can be
considered as relocalized (according to the terminology of
\cite{Van03a1}), because they can introduce nonlocality through a
generalized entropy current. A different continuum granular model of
Aranson and Tsimring introduces also a scalar internal variable with a
Ginzburg-Landau dynamics (assumed in an ad-hoc way)
\cite{AraTsi01m,Sapata02m,VolAta03m}. Their starting point is the
granular kinetic theory and thorough calculations and simulations try
to understand the connections in that direction. Therefore they model
does not compatible to a realistic static case. In understanding the
granular phenomena and especially to create an applicable continuum
model is a great challenge of physics where definitely a large
extension of our understanding of continuum concepts and therefore the
extension of continuum theories is necessary \cite{Cap03a}. The simple
model shown in this work is hopefully is a step in that direction
demonstrating the capabilities and simplicity of thorough constitutive
reasoning.

Finally, it is important to observe the differences in the conditions
of the derivation and also in the final results of the model of
Goodman-Cowin and the present one. Here the fluid like state space and
the large differences in the compressibility were essential conditions
in the derivation. There was no need to introduce a balance of
substructural interactions (but of course the state space can be
extended to include $\ddot{\gamma}$ and one can investigate the
consequences). We have got a Coulomb-Mohr failure criteria in the
static limit in both cases. However,the structure of the equilibrium
pressure tensors and therefore the implied physical reasons were
completely different. To investigate the physical relevance of the two
models further work is necessary. In this direction are very important
the known solutions and generalization of the Goodman-Cowin model like
the flow calculations of \cite{WanHut99a,Mas01a}. Another possibility
to test the differences could be to look into static failure data of
Coulomb-Mohr materials. Related experimental data regarding porous
rocks is published recently by V\'as\'arhelyi
\cite{Vas03a,Vas03a2,Vas02p}

\section{Acknowledgements}

I gratefully thank Paolo Maria Mariano for his valuable remarks. This
research was supported by OTKA T034715 and T034603.

\end{document}